# Nearly Free Electron States in MXenes

Mohammad Khazaei,[1,*] Ahmad Ranjbar,[2,*] Mahdi Ghorbani-Asl,[3] Masao Arai,[1] Taizo Sasaki,[1] Yunye Liang,[4] and Seiji Yunoki[2,5,6]

[1]Computational Materials Science Unit, National Institute for Materials Science (NIMS), 1-1 Namiki, Tsukuba 305-0044, Ibaraki, Japan

[2]Computational Materials Science Research Team, RIKEN Advanced Institute for Computational Science (AICS), Kobe, Hyogo 650-0047, Japan

[3]Department of Materials Science and Metallurgy, University of Cambridge, 27 Charles Babbage Road, Cambridge, CB3 0FS, UK

[4]New Industry Creation Hatchery Center, Tohoku University, Sendai 980-8579, Japan

[5]Computational Condensed Matter Physics Laboratory, RIKEN, Wako, Saitama 351-0198, Japan

[6]Computational Quantum Matter Research Team, RIKEN Center for Emergent Matter Science (CEMS), Wako, Saitama 351-0198, Japan

*Corresponding authors: khazaei.mohammad@nims.go.jp, ranjbar@riken.jp

**ABSTRACT:** Using a set of first-principles calculations, we studied the electronic structures of two-dimensional transition metal carbides and nitrides, so called MXenes, functionalized with F, O, and OH. Our projected band structures and electron localization function analyses reveal the existence of nearly free electron (NFE) states in variety of MXenes. The NFE states are spatially located just outside the atomic structure of MXenes and are extended parallel to the surfaces. Moreover, we found that the OH-terminated MXenes offer the NFE states energetically close to the Fermi level. In particular, the NFE states in some of the OH-terminated MXenes, such as $Ti_2C(OH)_2$, $Zr_2C(OH)_2$, $Zr_2N(OH)_2$, $Hf_2C(OH)_2$, $Hf_2N(OH)_2$, $Nb_2C(OH)_2$, and $Ta_2C(OH)_2$, are partially occupied. This is in remarkable contrast to graphene, graphane, and $MoS_2$, in which their NFE states are located far above the Fermi level and thus they are unoccupied. As a prototype of such systems, we investigated the electron transport properties of $Hf_2C(OH)_2$ and found that the NFE states in $Hf_2C(OH)_2$ provide almost perfect transmission channels without nuclear scattering for electron transport. Our results indicate that these systems might find applications in nanoelectronic devices. Our findings provide new insights into the unique electronic band structures of MXenes.

## Introduction:

Nearly free electron (NFE) states are those, which exhibit parabolic energy dispersions with respect to the crystal wave vector. The signature of such states can be found experimentally using angle-resolved photoemission spectroscopy and theoretically using band structure calculations [1]. The NFE states have been observed in the surface states of $C_{60}$ and $C_6F_6$ islands on copper surfaces [2,3], and predicted in the surface states of carbon and boron nitride nanotube bundles [4-7], in graphite [8-10], graphene [11], graphene nanoribbons [12], graphane [13], $MoS_2$ [14], and $Ca_2N$ [15]. It has been anticipated that the NFE states can be ideal electron transport channels without nuclear scattering [15]. In addition, the existence of such trapped states on the surface might promote the catalytic properties [16]. Also, the occupied NFE bands play a crucial role in high temperature superconductivity of alkali metals intercalated graphite compounds to facilitate strong electron-phonon coupling [17,18]. The NFE states are usually unoccupied and exist at high energies near the vacuum level, which makes them unsuitable for electron transport or other applications.[6] However, it has been shown that the empty NFE states might possibly be occupied and shifted toward the Fermi level by carrier injection and external electric field [6-10].

Recently a novel family of two-dimensional transition metal carbides and nitrides, so-called MXenes [19-22], has been synthesized from the exfoliation process of MAX phases using hydrofluoric acid [19,20]. The MAX phases are a family of layered compounds with chemical formula of $M_nAX_{n+1}$ (n=1-3), where M, A, and X are an early transition metal, an element from groups 13-16 in the periodic table, and carbon/nitrogen, respectively [23,24]. During the exfoliation process, the A element is washed out and the surfaces of the MXenes are terminated by the mixture of F, O, and OH groups [19,20]. Very recently, a significant advance in the growth of high quality crystalline MXenes has been achieved by chemical vapor deposition technique [25]. Moreover, the family of MXenes has been expanded even to double transition metals carbides $M'_2M''C_2$ and $M'_2M''_2C_3$ [26]. The MXenes have already found applications in Li ion batteries and high volumetric capacitors [27-32], transparent conductive electrodes [33], in electronic devices with high electron mobilities [34], filler in polymeric composites [35], purifier [36], and suitable substrates for dyes [37]. Theoretically, many applications were proposed for the MXenes in the electronic [38-44], magnetic [45], and thermoelectric [38,46], devices. Some of the MXenes are topological insulators [47,48]. MXenes can be used as sensors [49], ultralow work function materials [50], and Schottky barrier junctions [51].

In this work, using a set of first-principles calculations, we study the surface states of $M_2C$ (M= Sc, Ti, Zr, Hf, V, Nb, Ta) and $M'_2N$ (M'=Ti, Zr, Hf) functionalized with F, OH and O. It is found that some of the functionalized MXenes exhibit the NFE states, which are spatially localized above the surfaces. In contrast to the most of the available 2D systems such as graphene, graphane, boron nitride, $MoS_2$ [11,13,14], the NFE states in several OH-terminated MXenes are located near the Fermi level, and thus they may play a significant role in electron transport in nanoelectronic devices, catalysts, or gas sensors.

## Calculation method:

The first-principles electronic structure calculations were carried out based on generalized gradient approximation (GGA) in density functional theory with Perdew-Burke-Ernzerhof (PBE) [52] for exchange correlation function as implemented in VASP code [53]. The atomic positions and cell parameters were optimized using conjugate gradient method without imposing any symmetry. To avoid any artificial interaction between the layers and their images, a large vacuum space of 40 Å was set along the direction perpendicular to the surface. In the optimized structures, the maximum force on each atom is less than 0.005 eV/Å. The total energies were converged within $10^{-6}$ eV/cell. For the structural optimization, the Brillouin zone was sampled using a set of 12×12×1 $k$ points [54]. The Methfessel–Paxton smearing scheme was applied with a smearing width of 0.1 eV [55].

Electron localization function (ELF) was used to describe the chemical bonds [56,57], which is defined as $ELF = 1/(1 + (D/D_h))$, where $D = \frac{1}{2}\sum_i |\nabla\varphi_i|^2 - \frac{1}{8}\frac{|\nabla\rho|^2}{\rho}$ and $D_h =$ $\frac{3}{10}(3\pi^2\rho)^{5/3}$. The $\varphi_i$ represents the Kohn Sham orbitals and $\rho = \sum_i |\varphi_i|^2$ stands for electron charge density. ELF provides a value between 0 and 1. ELF = 1 corresponds to perfect localization of electrons and 0.5 indicates a uniform electron gas [56,57].

The coherent transport calculations were performed via non-equilibrium Green's function (NEGF) scheme as implemented in the Atomistix ToolKit (ATK) [58]. The calculations were done using GGA/PBE functional with double-zeta polarization numerical basis sets. We used an energy cutoff of 500 Ry and 0.00001 Ry of tolerance for the self-consistent calculations. The Brillouin zone of the transport model was sampled with 1×4×100 *k*-points. The current through the device was obtained at finite bias voltages using the Landauer formula and the nonequilibrium Green's function method [59]. The Poisson-Schrödinger equation of the system was self-consistently solved using a fast Fourier transform solver. The Landauer formula and the nonequilibrium Green's function method has already been applied to low-dimensional systems for the electron transport, e.g., see Refs. [60-63].

## Results and discussion:

### A. Existence and origin of the NFE states

Figure 1 shows projected band structures of the OH-terminated MXenes. The color and size of dots represent the projected weights of the wave functions onto different composed elements. We have also calculated the projected band structures of the O- and F-terminated MXenes (see Figs. S1 and S2 [64]). The band structure calculations indicate that $Sc_2CF_2$, $Sc_2C(OH)_2$, $Sc_2CO_2$, $Ti_2CO_2$, $Zr_2CO_2$, and $Hf_2CO_2$ are semiconductors with band gaps of 1.03, 0.45, 1.8, 0.24, 0.88, and 1.0 eV, respectively. In this study, all semiconducting MXenes except $Sc_2C(OH)_2$ have indirect band gaps [38].

In all MXenes studied here, the transition metals have dominant contributions to most of the states around the Fermi level. However, it is surprisingly to notice that in the OH-terminated MXenes some of the states around Γ point, near and above the Fermi level, do not constitute any of the composed elements. As will be shown later, these states are mainly located in vacuum region outside of the hydrogen atoms. They have NFE characteristics with parabolic energy dispersions with respect to the crystal wave vector. More interestingly some of the NFE states in $Ti_2C(OH)_2$, $Zr_2C(OH)_2$, $Zr_2N(OH)_2$, $Hf_2C(OH)_2$, $Hf_2N(OH)_2$, $Nb_2C(OH)_2$, and $Ta_2C(OH)_2$ are partially occupied. The NFE states of $Sc_2C(OH)_2$, $Ti_2N(OH)_2$, and $V_2C(OH)_2$ are unoccupied and located above the Fermi level, but in much lower energies than those of the graphene, graphane, BN, and $MoS_2$ (see Fig. S3 [64]). Therefore, the unoccupied NFE states in OH-terminated MXenes should be more accessible than those in other known 2D systems. In the F- and O-terminated MXenes, the NFE states are also observed, but located at high energies above 4.0 eV (see Figs. S1 and S2 [64]). This implies that the usage of the NFE states of F- and O-terminated MXenes will be challenging to employ for any application purposes.

Visualization of the ELF [56,57] is a quick way to track the existence of the partially occupied NFE states, which are spatially confined in narrow regions on top of the surfaces [15]. Figure 2(a) shows the ELF for $Hf_2C(OH)_2$. We found that a uniform floating electron gas with ELF ≈ 0.4 exist above the hydrogen atoms. We also found that all the OH-terminated MXenes where their NFE states are partially occupied show similar ELF patterns.

It has been extensively studied in the literature that the tails of local electrostatic potential beyond the surfaces, which are mostly formed by adsorbate overlayers and electrostatic image potentials, induce series of bound states near or above the Fermi level [13,65-69]. It has also been shown that the energetic of the NFE states are sensitive to the surface dipole moments [70]. Margine and Crespi performed a detailed study of the effect of positive surface charges on energetic of NFE states in alkali deposited carbon nanotubes by using the jellium model [71]. They could show clearly that the downshift arises from an electrostatic mechanism: the NFE states concentrate in the region of highest positive charges. It was shown that as the surface positive charge density increases, the NFE states get closer to the Fermi level. Also it was indicated that at a certain positive charge density, the NFE states drop so rapidly in energy relative to the other conduction bands and cross the Fermi level. Moreover, it was shown that the NFE states that are already located below the Fermi level do not shift significantly to deeper energies and are basically pinned near the Fermi level, even if higher positive charge density is deposited on the surface. This might be due to screening effect or hybridization of the NFE states with other occupied states [72]. The same physics is applicable for the OH-MXenes, in which the H atoms are positively charged. Hence, the NFE states appear near the Fermi level in OH-terminated MXenes.

In first-principles calculations, the estimation of surface charges is not straightforward. However, the effect of surfaces charges on electrostatic potential in vacuum can be obtained accurately. Hence, in order to better understand why the NFE states in the OH terminated MXenes are found near the Fermi level, but not in the F- and O-terminated MXenes, we investigated the average effective total potential for $Hf_2C(OH)_2$, $Hf_2CF_2$ and $Hf_2CO_2$ along the $c$-axis perpendicular to their surfaces, and the results are shown in Figs. 2(b)-(d). Here, the calculated total potential includes ionic, Hartree, and exchange-correlation terms. It is observed in Figs. 2(b)-(d) that the tail of potential above the H are shallower and spatially more extended in $Hf_2C(OH)_2$ as compared to the cases for $Hf_2CF_2$ and $Hf_2CO_2$. The NFE states can be only formed in the spatial regions away beyond the surfaces toward the vacuum in the electrostatic potential wells. In this regard, the shallow potential tails in the OH-terminated MXenes is found near the Fermi level, while the shallow potential tails in the F- and O-terminated systems are located far above the Fermi level. Hence, the NFE states of the OH-terminated MXenes are generally found near the Fermi level, while those in F- and O-terminated MXenes appear at higher energies. To demonstrate the spatial localization, the average charge density of the lowest NFE state for the $Hf_2C(OH)_2$ at Γ point is shown in Fig. 2(b). It is clearly observed that this NFE state is mostly localized in a spatially narrow region above the hydrogen atoms and exponentially decays toward the vacuum.

In order to understand qualitatively the systematic variation of the position of the NFE states and the vacuum level, we have summarized in Fig. 3 the relative energy position of the lowest NFE states with respect to the Fermi level for F-, O-, and OH-terminated MXens as well as for graphene, graphane, BN, and $MoS_2$ layers as a function of the work function. The work function is a material specific parameter that quantifies the energy difference between the Fermi level and the vacuum level. The work function basically indicates the minimum electrostatic energy required to remove a single electron away from the surface. The energetic of the NFE states and the work function parameters are interrelated through the potential energy surface. It is observed in Fig. 3 that the NFE states appear just below the vacuum level in neutral charge surfaces such as graphene, graphane, BN and $MoS_2$ as well as in the negatively charged surfaces such as F- and O-terminated MXenes. However, the NFE states in the OH-terminated MXenes with positively charged surfaces are found near the Fermi level. Note that the low work function properties of the OH-terminated MXenes as compared with the F- and O-terminated MXenes can be attributed to the dipole moments generated by the OH groups [50].

As seen from Fig. 3, all NFE states are far below the vacuum level in OH functionalized MXenes. This means that the energy pinning has already happened and the positive surface charging is strong. In order to understand why the lowest NFE states in some of the OH terminated MXenes locate above the Fermi level (see Fig. 3), the analysis of the electronic structures of MXenes may be necessary. The largest deviation of energetic of the lowest NFE state from the Fermi level (0.24 eV) belongs to $Sc_2C(OH)_2$, which is a semiconducting system. Our results indicate that for the semiconducting systems, a stronger positive surface charging might be necessary to shift the NFE states to below Fermi level. To explain why in some of the metallic OH-terminated MXenes the lowest NFE state locate slightly above the Fermi level < 0.062 eV, it was appropriate if we could analyze the amount of the surface charges quantitatively. However, unfortunately, as it is well known, there is no accurate method to estimate the charges. The available methods such as Mulliken or Bader charge analyses can be used only to see qualitatively that the OH terminated MXenes are positively charged systems, but cannot be used for accurate quantitative comparisons. Hence, the report of charge data cannot help us to explain the subtle differences in the position of NFE states in OH-terminated systems. Moreover, the subtle differences in the position of lowest NFE states in OH-terminated systems might be affected by the distances between the hydrogen atoms on the surfaces, which are related to lattice constants of the MXenes. In this regard, in the previous study on functionalized graphene and polysilane, it has been predicted that a smoother surface is expected to stabilize NFE states [70]. However, more detailed analysis is necessary in the future work.

### B. Robustness of the NFE states

We studied the robustness of the NFE states against various modifications by thickness, strain, gas adsorption and formation of heterostructure. The band structures and the ELF calculated for $Hf_{n+1}C_n(OH)_2$ with n=1-3 were calculated (see Fig. S4 [64]). We found the existence of the NFE states and realized that the location of these NFE energy bands relative to the Fermi level do not depend significantly on the thickness n of the MXenes. This is simply because, as explained above, the NFE characteristics are basically determined by the shape of the electrostatic potential outside the hydrogen atoms and the energy of the potential tail relative to the Fermi level, which are both insensitive to the thickness n of MXenes. In addition, the nature of surfaces charges of the MXenes does not depend on the thickness of the MXenes, *i.e.*, independent of the thickness, the surfaces of the OH terminated (F and O-terminated) MXenes are positively (negatively) charged.

We also investigated the robustness of the occupied NFE states found in $Hf_2C(OH)_2$ under external strains quantified by $\Delta a/a$, where $a$ is the optimized lattice parameter and $\Delta a$ is the change of the lattice constant from $a$ (see Figs. S5 and S6 [64]). From the band structure and ELF calculations it is found that the partially occupied NFE states exist just outside the surfaces at strains between -4% and +4%. However, with further increasing compressive or tensile strains, the partially occupied NFE states near the Fermi level are eventually disappeared.

Moreover, we examined the interaction of $H_2$, $O_2$, $N_2$, CO, $NH_3$, and $CH_4$ molecules with $Hf_2C(OH)_2$ considering a $3\times3\times1$ supercell (see Fig. S7 [64]). We found that the interaction between the NFE states and some of the adsorbed molecules such as $O_2$, $N_2$, and CO are strong that the NFE states are completely disappeared even at low molecule adsorption concentrations. This implies that the OH-terminated MXenes with the occupied NFE states are highly sensitive to gas adsorbents. Thereby, they are possibly applied for selective gas sensor applications. Indeed, semiconducting $Ti_2CO_2$ has been suggested for the selective $NH_3$ sensing [49]. Also, it has been proposed that due to their localized states that appear at above the hydrogen atoms the OH-terminated MXenes may offer potential application toward Pb or heavy metal purification for environmental remediation [36,73]. Similar to our observation, a previous study on the NFE states on $Ca_2N$ found that the absorption of the molecules could destroy the NFE states. It should be noted that the NFE states of $Hf_2C(OH)_2$ overlap with other states at the Fermi level, but the NFE state of $Ca_2N$ monolayer is the only state at the Fermi level [15]. Indeed, $Ca_2N$ belongs to a set of materials known as 2D-electrides, in which electrons near the Fermi level float between the ionic layers [74]. It was shown that a graphane sheet, but not graphene and BN sheets, could protect the NFE states of the $Ca_2N$ [15]. The above study motivated us to study the electronic structure of $Hf_2C(OH)_2$ when it is placed on a single layered graphene, as will be explained in the following.

These studies motivated us to examine the electronic structure of $Hf_2C(OH)_2$ when it is placed on a single layered graphene. In this regard, in order to minimize the lattice mismatch, we considered heterostructure of graphene/$Hf_2C(OH)_2$ composed of a 3×3×1 supercell of $Hf_2C(OH)_2$ and a 4×4×1 supercell of graphene, for which the lattice mismatch is less than 0.4%. The projected density of states, band structure, and ELF for the graphene/$Hf_2C(OH)_2$ heterostructure are shown in Figs. 4(a)-(c). The band structure and ELF calculations show that there is no NFE state at the interface of the heterostructre. Because of the small distance between hydrogen atoms and graphene (2.27 Å), the local potential above the hydrogen atoms is vanished. Thereby, electrons cannot be trapped at the interface. Instead, they are localized either on $Hf_2C(OH)_2$ or on graphene. It is however clear in Fig. 4(c) that the NFE state remains stable on the other side of $Hf_2C(OH)_2$ without graphene. The ELF analysis indicates that there is no chemical bond between graphene and the $Hf_2C(OH)_2$. The $Hf_2C(OH)_2$ is adsorbed on graphene physically with binding energy of -0.11 eV/graphene cell. We also investigated the projected band structure and ELF for the case where both sides of the $Hf_2C(OH)_2$ are covered by graphene layers, and found that the NFE states, which exist in $Hf_2C(OH)_2$ are all disappeared from the Fermi level. Moreover, we performed similar calculations on BN/$Hf_2C(OH)_2$/BN and graphane/$Hf_2C(OH)_2$/graphane heterostructures. Although BN and graphane are large gap insulators, they cannot protect the partially occupied NFE states. In agreement with our results, the previous study on $Ca_2N$ [15] found that graphene or BN could not protect the NFE states. However, in contrast with our results, it was suggested that graphane with a large band gap can protect the NFE state on $Ca_2N$. The ineffectiveness protection of the graphane for the NFE states on $Hf_2C(OH)_2$ is understood because both $Hf_2C(OH)_2$ and graphane have hydrogen atoms on their surfaces. Therefore, due to hydrogen bonding effect, the layers of graphane and $Hf_2C(OH)_2$ tend to get closer, which results in destroying of the NFE states. One possible way to protect the NFE states in OH-terminated MXenes is to fabricate their nanotubes or conical scrolls. Since in the 2D $Hf_2C(OH)_2$ monolayer, the NFE states can be found on both surfaces, by rolling this monolayer, the NFE of inner surface still exists. In this regards, it is noteworthy that conical scroll of MXenes exist experimentally [19].

In addition, we found in Figs. 4(a) and (b) that graphene is n-doped by $Hf_2C(OH)_2$ due to the low work function property of the $Hf_2C(OH)_2$ (2.32 eV) in comparison with graphene (4.26 eV). However, the Dirac-cone like energy dispersion of graphene remains intact, and is simply shifted to lower energies at -1.1 eV by receiving electrons from $Hf_2C(OH)_2$. In order to depict the charge transfer between $Hf_2C(OH)_2$ and graphene clearly, we calculated in Fig. 4(d) $\Delta\rho=\rho-[\rho_{graphene}+\rho_{Hf2C(OH)2}]$, where $\rho$, $\rho_{graphene}$, and $\rho_{Hf2C(OH)2}$ are the total charge densities of the graphene/$Hf_2C(OH)_2$ heterostructure, graphene, and $Hf_2C(OH)_2$, respectively. Here, $\rho_{graphene}$ ($\rho_{Hf2C(OH)2}$) is generated from a total energy calculation of the graphene ($Hf_2C(OH)_2$) by removing the $Hf_2C(OH)_2$ (graphene) from the heterostructure without further optimizing the atomic structures. As shown in Fig. 4(d), the charges are mainly transferred from OH bonds of the $Hf_2C(OH)_2$ to $\pi^*$ states of the graphene. Because of the semi-metallic and metallic natures of graphene and $Hf_2C(OH)_2$, respectively, it is expected that the graphane/$Hf_2C(OH)_2$ heterostructure exhibits metallic current-voltage characteristics.

We also examined the electronic structures of multilayer systems where the OH-terminated MXenes are stacked periodically (see Fig. S8 [64]). We found within GGA/PBE calculations that the multilayer systems, which are constructed from the MXenes with partially occupied NFE states possess the partially occupied NFE states as long as the interlayer distance is relatively large > 3.28 Å. However, when the van der Waals interactions are semi-empirically included with DFT-D2 method [75], we found that the partially occupied NFE states disappear (see Fig. S8 [64]). This is because the van der Waals interactions decrease the interlayer distances significantly (~1.96 Å), and thus the tails of confinement potentials vanish between the hydrogen atoms in the consecutive MXene layers. Therefore, in order to stabilize the partially occupied NFE states in the multilayered OH-terminated MXenes, the hydrogen atoms belonging to any two consecutive layers should be kept away from each other. This should be possible by intercalating inert dopants or molecules between the layers with low concentrations.

Although experimentally several MXenes, such as $Ti_2C$, $V_2C$, $Nb_2C$, $Ta_2C$, $Mo_2C$, TiNbC, $(Ti_{0.5}V_{0.5})_2C$, $Ti_3C_2$, $Ti_3CN$, $(Nb_{0.8}Ti_{0.2})_4C_3$, and $(Nb_{0.8}Zr_{0.2})_4C_3$ have already been synthesized with mixture of F, OH, and O surface functionalization [19-22,27,29], the preparation of MXenes with a specific surface functionalization is still challenging and needs to be pursued. However, we have previously studied phonon dispersions for various MXenes [50] and shown that $Sc_2C(OH)_2$, $Ti_2C(OH)_2$, $Zr_2C(OH)_2$, $Zr_2N(OH)_2$, $Hf_2C(OH)_2$, $Hf_2N(OH)_2$, and $Ta_2C(OH)_2$ are dynamically stable (see Fig. S9 [64]). Therefore, the OH terminated MXenes will be experimentally realized in the near future.

Some of the MXenes such as $V_2C$, $Nb_2C$, and $Ti_3C_2$ have been experimentally applied for Li, Na, or Ca ion batteries [29,76]. Furthermore the previous theoretical studies have shown that hydrogen atoms can be removed from the OH-terminated MXenes upon alkali metal adsorption to form $M_{n+1}C_nO_2A_2$ (A = Li, Na, Ca) [30]. Motivated by these studies, we have also calculated the projected band structures for $M_{n+1}C_nO_2Li_2$ (M =Sc, Ti, V, Zr, Nb, Hf, and Ta) and $M'_{n+1}N_nO_2Li_2$ (M' = Ti, Zr, and Hf) monolayers (see Fig. S10 [64]). We have also found the NFE states near the Fermi level in $M_{n+1}C_nO_2Li_2$ and $M'_{n+1}N_nO_2Li_2$ MXenes. This is because the Li has an electropositive nature similar to H in OH-terminated MXenes.

## C. Electron transport properties

Motivated by the fascinating electronic structure of the OH-terminated MXenes, we also examined the quantum electron transport for $Hf_2C(OH)_2$ monolayer, which exhibits two partially occupied NFE states near the Fermi level. The transport simulation setup consists of the MXene in a scattering region and two semi-infinite electrodes that are also made of the MXene, as shown in Fig. 5(a).

As we explained above, the ATK code uses a localized basis set to describe the wave functions with the number of bases being smaller than the number of plane wave bases used in VASP code. Though the short-range localized basis set used in the ATK code makes performing the first-principles calculations faster, it might have a drawback on the correct representation of the NFE states, which are mainly located outside the atomic network of the MXenes. Hence, before performing any transport calculation using ATK, we examined the accuracy of the ATK basis set to capture the NFE band features of $Hf_2C(OH)_2$ near the Fermi found in the VASP calculations. The results of the band structure calculations were compared (see Fig. S11 [64]). It is observed that the NFE band features calculated by ATK are slightly different from those obtained by VASP, especially near the Fermi level. In order to better reproduce the NFE band feature, we then included ghost atoms above the H atoms. The band structure of $Hf_2C(OH)_2$ was calculated with ghost atoms (see Fig. S11 [64]). The ghost atom used in ATK is an atom with a localized basis set but without any pseudopotential core or charge [58]. However, the local orbitals of the ghost atoms can be populated in order to host a finite electron density in a region where there are no real atoms. The alternative to having a ghost atom on the surface is to extend the range of the basis functions, which increases the computational cost [58]. Thereby, we adopted the ghost atom scheme for the electron transport calculations.

Figure 5(b) shows the energy-dependent transmission function of the $Hf_2C(OH)_2$ at different bias voltages. For more detail, the transmission curve (the k-point averaged transmission spectrum as function of energy) for each bias voltage is separately shown in Fig. S12. In Fig. 5(b) [64], the white dotted lines indicate the range of the integration window of the transmission that is used to calculate the conducting current through the scattering region at each bias voltage. The transmission spectrum shows an asymmetric behavior with larger contribution of electrons than holes in the transport within the energy integration window. With increasing the bias voltage, the transmission peaks for electrons (holes) are shifted to higher (lower) energies and their heights are suppressed. The shifts of transmission peaks have been highlighted in Fig. 5(b) by white arrows. As the bias voltage increases, the integration window of the transmission increases, and consequently the current increases almost linearly with the applied voltage as shown in Fig. 5(c).

In order to visualize the signature of the NFE states in the electron transport, we have calculated the eigenvalues and eigenvectors of the transmission matrix at a given energy value, kpoint, and a given bias voltage. The eigenvalue corresponds to the magnitude of the transmission probability --- which lies in the interval of 0 and 1 --- for each transport channel through the scattering region, described by the corresponding eigenvector. As an example, we calculated in Fig. 6 the eigenvalues and eigenvectors of the transmission matrix at a typical energy of -0.088 eV in which the occupied NFE states exist at zero bias voltage. As shown in Fig. 6, the transmission matrix has five eigenvalues, and consequently five transport channels. The eigenvectors of n=3 and 4 can be considered as NFE transport channels because their issosurfaces locate mainly at above the hydrogen atoms. In contrast, the eigenvectors of n=0, 1, and 2 cannot be considered as NFE transport channels because they are either localized on the MXene surfaces (n=0 and 1) or have a node (n=2).

In addition, we also investigated the effect of bias voltages on the existence of the NFE states and found that the number of hole transmission channels (i.e., the transmission channels below Fermi level) decreases with increasing the bias voltage. As an example, we calculated the eigenvalues and eigenvector of transmission matrix at -0.088 eV under various bias voltages. The transmission probability of the partially occupied NFE decreases upon applying bias voltages and are vanished at bias voltage > 0.6 V (see Fig. S13 [64]). However in contrast, the NFE states that were at the Fermi level are shifted to higher energies without significant change in their transmission probabilities (see Fig. S14 [64]). Our transmission analysis clearly show that the NFE states can act as both hole as well as conduction channels at low bias voltages, thus suitable for applications in low-power nanoelectronic devices. Moreover, considering their low work functions [50], the OH-terminated MXenes with partially occupied NFE states might be applied for excellent electron field emitters.

## Conclusion:

We have studied the surface electronic structures of $M_2C$ (M= Sc, Ti, Zr, Hf, V, Nb, Ta) and $M'_2N$ (M'=Ti, Zr, Hf) functionalized with F, OH and O. We have found that the OH-terminated MXenes offer the NFE states near the Fermi level. Our calculations indicate that the partially occupied NFE states of the OH-terminated MXenes vanish upon the surface coverage by other 2D systems such as graphene, BN, or graphane. In order to stabilize the partially occupied NFE states in the multilayered OH-terminated MXenes, the interlayer H…H distances should be larger than ~3.3 Å. Experimentally, such systems might be feasible by intercalation of inert molecules at low concentrations without significantly deforming the local potentials outside the hydrogen atoms in the consecutive MXene layers. Furthermore, our transport calculations show that the NFE states contribute to the electronic transport free of nuclear scattering. Since the NFE states locate above the MXene surfaces away from the hydrogen atoms, the electrons can conduct through the NFE channels without being significantly scattered by the hydrogen vibrations at finite temperatures. Our predictions on the NFE states in MXenes should stimulate further studies on deliberate surface state engineering of MXenes for electronic, catalytic, and sensor applications.

ACKNOWLEDGMENT

The calculations were performed on Numerical Materials Simulator at National Institute for Materials Science (NIMS) and also partially on RIKEN supercomputer system (HOKUSAI GreatWave) and the HITACHI SR16000/M1 supercomputer of the Institute for Materials Research, Tohoku University.

REFERENCES

[1] N. Alisdoust, G. Bian, S. −Y. Xu, R. Sankar, M. Neupane, C. Liu, L. Belopolski, D. −X. Qu, J. D. Denlinger, F. −C. Chou, and M. Z. Hasan, Observation of monolayer valence band spin-orbit effect and induced quantum well states in $MoX_2$, Nat. Commun. **5**, 4673 (2014).

[2] M. Feng, J. Zhao, T. Huang, X. Zhu, and H. Petek, The electronic properties of superatom states of hollow molecules, Acc. Chem. Res. **44**, 360 (2011).

[3] D. B. Dougherty, M. Feng, H. Petek, J. T. Yates, Jr., and J. Zhao, Band formation in a molecular quantum well via 2D superatom orbital interactions, J. Phys. Rev. Lett. **109**, 266802 (2012).

[4] S. Okada, A. Oshiyama, and S. Saito, Nearly free electron states in carbon nanotube bundles, Phys. Rev. B **62**, 7634 (2000).

[5] S. Hu, J. Zhao, Y. Jin, J. Yang, H. Petek, and J. G. Hou, Nearly free electron superatom states of carbon and boron nitride nanotubes, Nano Lett. **10**, 4830 (2010).

[6] J. Zhao, Q. Zheng, H. Petek, and J. Yang, Nonnuclear nearly free electron conduction channels induced by doping in nanotube-molecular sheet composites, J. Phys. Chem. A **118**, 7255 (2014).

[7] B. Yan, C. Park, J. Ihm, G. Zhou, W. Duan, and N. Park, Electron emission originated from free-electron-like states of alkali-doped boron-nitride nanotubes, J. Am. Chem. Soc. **130**, 17012 (2008).

[8] M. Posternak, A. Baldereschi, A. J. Freeman, and E. Wimmer, Prediction of electronic surface states in layered materials: graphite, Phys. Rev. Lett. **52**, 863 (1984).

[9] M. Otani and S. Okada, Field-induced free-electron carriers in graphite, J. Phys. Soc. Jpn. **79**, 073701 (2010).

[10] M. Otani and S. Okada, Gate-controlled carrier injection hexagonal boron nitride, Phys. Rev. B **83**, 073405 (2011).

[11] V. M. Silkin, J. Zhao, F. Guinea, E. V. Chulkov, P. M. Echenique, and H. Petek, Image potential states in graphene, Phys. Rev. B **80**, 121408 (2009).

[12] A. Yamanaka and S. Okada, Electron injection into nearly free electron states of graphene nanoribbons under a lateral electric field, Appl. Phys. Express **7**, 125 (2014).

[13] J. Zhao and H. Petek, Non-nuclear electron transport channels in hollow molecules, Phys. Rev. B **90**, 075412 (2014).

[14] N. T. Cuong, M. Otani, and S. Okada, Gate-induced electron-state tuning of $MoS_2$: first-principles calculations, J. Phys.: Condens. Matter **26**, 135001 (2014).

[15] S. Zhao, Z. Li, and J. Yang, Obtaining two-dimensional electron gas in free space without resorting to electron doping: an electride based design, J. Am. Chem. Soc. **136**, 13313 (2014).

[16] K. Nagesha and L. Sanche, Effect of band structure on electron attachment to adsorbed molecules: cross section enhancements via coupling to image states, Phys. Rev. Lett. **81**, 5892 (1998).

[17] G. Csányi, P. B. Littlewood, A. H. Nevidomskyy, C. J. Pickard, and B. D. Simons, The role of interlayer state in the electronic structure of superconducting graphite intercalated compounds, Nat. Phys. **1**, 42 (2005).

[18] M. Calandra and F. Mauri, Theoretical explanation of superconductivity in $C_6Ca$, Phys. Rev. Lett. **95**, 237002 (2005).

[19] M. Naguib, M. Kurtoglu, V. Presser, J. Lu, J. Niu, M. Heon, L. Hultman, Y. Gogotsi, and M. W. Barsoum, Two-Dimensional nanocrystals produced by exfoliation of $Ti_3AlC_2$, Adv. Mater. **23**, 4248 (2011).

[20] M. Naguib, O. Mashtalir, J. Carle, V. Presser, J. Lu, L. Hultman, Y. Gogotsi, and M. W. Barsoum, Two-dimensional transition metal carbides, ACS Nano **6**, 1322 (2012).

[21] R. Meshkian, L. –Å. Näslund, J. Halim, J. Lu, M. W. Barsoum, and J. Rosen, Synthesis of two-dimensional molybdenum carbide, $Mo_2C$, from the gallium based atomic laminate $Mo_2Ga_2C$, Scripta Mater. **108**, 147 (2015).

[22] J. Yang, M. Naguib, M. Ghidiu, L. –M. Pan, J. Gu, J. Nanda, J. Halim, Y. Gogotsi, and M. W. Barsoum, Two-dimensional Nb-based $M_4C_3$ solid solutions (MXenes), J. Am. Ceram. Soc. **99**, 660 (2015).

[23] M. Khazaei, M. Arai, T. Sasaki, M. Estili, and Y. Sakka, Trends in electronic structures and structural properties of MAX phases: a first-principles study on $M_2AlC$ (M = Sc, Ti, Cr, Zr, Nb, Mo, Hf, or Ta), $M_2AlN$, and hypothetical $M_2AlB$ phases, J. Phys.: Condens. Matter **26**, 505503 (2014).

[24] M. Khazaei, M. Arai, T. Sasaki, M. Estili, and Y. Sakka, The effect of the interlayer element on the exfoliation of layered $Mo_2AC$ (A = Al, Si, P, Ga, Ge, As or In) MAX phases into two-dimensional $Mo_2C$ nanosheets, Sci. Tech. Adv. Mater. **15**, 014208 (2014).

[25] C. Xu, L. Wang, Z. Liu, L. Chen, J. Guo, N. Kang, X. –L. Ma, H. –M. Cheng, and W. Ren, Large-area high-quality 2D ultrathin $Mo_2C$ superconducting crystals, Nat. Mater. **14**, 1135 (2015).

[26] B. Anasori, Y. Xie, M. Beidaghi, J. Lu, B. C. Hosler, L. Hultman, P. R. C. Kent, Y. Gogotsi, and M. W. Barsoum, Two-dimensional, ordered, double transition metals carbides (MXenes), ACS Nano **9**, 9507 (2015).

[27] M. R. Lukatskaya, O. Mashtalir, C. E. Ren, Y. Dall'Agnese, P. Rozier, P. L. Taberna, M. Naguib, P. Simon, M. W. Barsoum, and Y. Gogotsi, Cation intercalation and high volumetric capacitance of two-dimensional titanium carbide, Science **341**, 1502 (2013).

[28] D. Sun, Q. Hu, J. Chen, X. Zhang, L. Wang, Q. Wu, and A. Zhou, Structural transformation of MXene ($V_2C$, $Cr_2C$, and $Ta_2C$) with O groups during lithiation: a first-principles investigation, ACS Appl. Mater. Interfaces **8**, 74 (2016).

[29] M. Naguib, J. Halim, J. Lu, K. M. Cook, L. Hultman, Y. Gogotsi, and M. W. Barsoum, New two-dimensional niobium and vanadium carbides as promising materials for Li-ion batteries, J. Am. Chem. Soc. **135**, 15966 (2013).

[30] Y. Xie, Y. Dall'Agnese, M. Naguib, Y. Gogotsi, M. W. Barsoum, H. L. Zhuang, and P. R. C. Kent, Prediction and characterization of MXene nanosheet anodes for non-lithium-ion batteries, ACS Nano **8**, 9606 (2014).

[31] H. Pan, Electronic properties and lithium storage capacities of two-dimensional transition-metal nitride monolayers, J. Mater. Chem. A **3**, 21486 (2015).

[32] J. Zhu, A. Chroneos, and U. Schwingenshlögl, Nb-based MXenes for Li-ion battery applications, Phys. Status Solidi PRL **9**, 726 (2015).

[33] J. Halim, M. Lukatskaya, K. M. Cook, J. Lu, C. R. Smith, L. Å. Näslund, S. J. May, L. Hultman, Y. Gogotsi, P. Eklund, and M. W. Barsoum, Transparent conductive two-dimensional titanium carbide epitaxial thin films, Chem. Mater. **26**, 2374 (2014).

[34] S. Lai, J. Jeon, S. K. Jang, J. Xu, Y. J. Choi, J. –H. Park, E. Hwang, and S. Lee, Surface group modification and carrier transport property of layered transition metal carbides ($Ti_2CT_x$, T: -OH, -F, and -O), Nanoscale **7**, 19390 (2015).

[35] X. Zhang, J. Xu, H. Wang, J. Zhang, H. Yan, B. Pan, J. Zhou, and Y. Xie, Ultrathin nanosheets of MAX phases with enhanced thermal and mechanical properties in polymeric compositions: $Ti_3Si_{0.75}Al_{0.25}C_2$. Angew, Chem. Int. Ed. **52**, 4361 (2013).

[36] Q. Peng, J. Guo, Q. Zhang, J. Xiang, B. Liu, A. Zhou, R. Liu, and Y. Tian, Unique lead adsorption behavior of activated hydroxyl group in two-dimensional titanium carbide, J. Am. Chem. Soc. **136**, 4113 (2014).

[37] O. Mashtalir, K. M. Cook, V. N. Mochalin, M. Crowe, M. W. Barsoum, and Y. Gogotsi, Dye Adsorption and decomposition on two-dimensional titanium carbide in aqueous media, J. Mater. Chem. A **2**, 14334 (2014).

[38] M. Khazaei, M. Arai, T. Sasaki, C. -Y. Chung, N. S. Venkataramanan, M. Estili, Y. Sakka, and Y. Kawazoe, Novel electronic and magnetic properties of two-dimensional transition metal carbides and nitrides, Adv. Funct. Mater. **23**, 2185 (2013).

[39] Y. Xie and P. R. C. Kent, Hybrid density functional study of structural and electronic properties of functionalized $Ti_{n+1}X_n$ (X=C, N) monolayers, Phys. Rev. B **87**, 235441 (2013).

[40] A. N. Enyashin and A. L. Ivanovskii, Structural and electronic properties and stability of MXenes $Ti_2C$ and $Ti_3C_2$ functionalized by methoxy groups, J. Phys. Chem. C **117**, 13637 (2013).

[41] S. Zhao, W. Kang, and J. Xue, MXene nanoribbons, J. Mater. Chem. C **3**, 879 (2015).

[42] Y. Lee, S. B. Cho, and Y. -C. Chung, Achieving a direct band gap in oxygen functionalized-monolayer scandium carbide by applying an electric field, Phys. Chem. Chem. Phys. **16**, 26273 (2014).

[43] X. Zhang, X. Zhao, D. Wu, Y. Jing, and Z. Zhou, High and anisotropic carrier mobility in experimentally possible $Ti_2CO_2$ (MXene) monolayer and nanoribbons, Nanoscale **7**, 16020 (2015).

[44] X. –H. Zha, K. Luo, Q. Li, Q. Huang, J. He, X. Wen, and S. Du, Role of the surface effect on the structural, electronic and mechanical properties of the carbide MXenes, Europhys. Lett. **111**, 26007 (2015).

[45] C. Si, J. Zhou, and Z. Sun, Half metallic ferromagnetism and surface functionalization-induced metal-insulator transition in graphene-like two-dimensional $Cr_2C$, ACS Appl. Mater. Interfaces **7**, 17510 (2015).

[46] M. Khazaei, M. Arai, T. Sasaki, M. Estili, and Y. Sakka, Two-dimensional molybdenum carbides: potential thermoelectric materials of the MXene family, Phys. Chem. Chem. Phys. **16**, 7841 (2014).

[47] H. Weng, A. Ranjbar, Y. Liang, Z. Song, M. Khazaei, S. Yunoki, M. Arai, Y. Kawazoe, Z. Fang, and X. Dai, Large-gap two-dimensional topological insulator in oxygen functionalized MXene, Phys. Rev. B **92**, 075436 (2015).

[48] H. Fashandi, V. Ivády, P. Eklund, A. Lloyd Spetz, M. I. Katsnelson, and I. A. Abrikosov, Dirac points with giant spin-orbit splitting in the electronic structure of two-dimensional transition metal carbides, Phys. Rev. B **92**, 155142 (2015).

[49] X. -F. Yu, Y. Li, J. -B. Cheng, Z. -B. Liu, Q. -Z. Li, W. -Z. Li, X. Yang, and B. Xiao, Monolayer $Ti_2CO_2$: A promising candidate for $NH_3$ sensor or capturer with high sensitivity and selectivity, ACS Appl. Mater. Interfaces **7**, 13707 (2015).

[50] M. Khazaei, M. Arai, T. Sasaki, A. Ranjbar, Y. Liang, and S. Yunoki, OH-terminated two-dimensional transition metal carbides and nitrides as ultralow work function materials, Phys. Rev. B **92**, 075411 (2015).

[51] Y. Lee, Y. Hwang, Y. C. Chung, Achieving type I, II, and III heterojunctions using functionalized MXene, ACS Appl. Mater. Interfaces **7**, 7163 (2015).

[52] J. P. Perdew, K. Burke, and M. Ernzerhof, Generalized gradient approximation made simple, Phys. Rev. Lett. **77**, 3865 (1996).

[53] G. Kresse and J. Furthmüller, Efficiency of ab-initio total energy calculations for metals and semiconductors using a plane-wave basis set, Comput. Mater. Sci. **6**, 15 (1996).

[54] H. J. Monkhorst and J. D. Pack, Special points for brillouin-zone integrations, Phys. Rev. B **13**, 5188 (1976).

[55] M. Methfessel and A. T. Paxton, High-precision sampling for brillouin-zone integration in metals, Phys. Rev. B **40**, 3616 (1989).

[56] B. Silvi and A. Savin, Classification of chemical bonds based on topological analysis of electron localization functions, Nature **371**, 683 (1994).

[57] A. Savin, R. Nesper, S. Wengert, and T. E. Fässler, ELF: the electron localization function, Angew. Chem. Int. Ed. Engl. **36**, 1808 (1997).

[58] QuantumWise Atomistix ToolKit (ATK), http://www. quantumwise.com.

[59] S. Datta, Quantum Transport: Atom to Transistor, Cambridge University Press, New York, USA 2005.

[60] M. Khazaei, S. U. Lee, F. Pichierri, and Y. Kawazoe, Designing nanogadgets by interconnecting carbon nanotubes with zinc layers, ACS Nano **2**, 939 (2008).

[61] M. Khazaei, S. U. Lee, F. Pichierri, and Y. Kawazoe, Electron transport through carbon nanotube intermolecular heterojunctions with peptide linkages, J. Phys. Chem. C **111**, 12175 (2007).

[62] M. Ghorbani-Asl, S. Borini, A. Kuc, T. Heine, Strain-dependent modulation of conductivity in single-layer transition-metal dichalcogenides, Phys. Rev. B **87**, 235434 (2013).

[63] M. Ghorbani-Asl, N. Zibouche, and M. Wahiduzzaman, A. F. Oliveira, A. Kuc, T. Heine, Electromechanics in $MoS_2$ and $WS_2$: nanotubes vs. monolayers, Sci. Rep. **3**, 2961 (2013).

[64] See SupplementalMaterial at http://link.aps.org/supplemental/ for projected band structures of F- and O-terminated MXenes as well as for graphene, graphane, BN, and $MoS_2$ layers; projected band structures and ELF of the $Hf_3C_2(OH)_2$, and $H_4C_3(OH)_2$, band structures and ELFs of $Hf_2C(OH)_2$ under strain; ELF of different adsorbed molecules on $Hf_2C(OH)_2$; effect of van der Waals interactions on band structure of $Hf_2C(OH)_2$; phonon spectra of OH-terminated MXenes; Projected band structure of Li-terminated MXenes; ATK band structure of $Hf_2C(OH)_2$ with ghost atoms; transmission curves at various voltages; effect of applied voltages on transmission probability of the NFE states.

[65] N. V. Smith, Phase analysis of image states and surface states associated with nearly-free-electron band gaps, Phys. Rev. B **32**, 3549 (1985)

[66] R. Fischer, Th. Fauster, and W. Steinmann, Three-dimensional localization of electrons on Ag islands, Phys. Rev. B **48**, 15496 (1993).

[67] Th. Fauster, Quantization of electronic states on metal surfaces, Appl. Phys. A **59**, 479 (1994).

[68] E. Bertel, One- and two-dimensional surface states on metals, Surf. Sci. **331**, 1136 (1995).

[69] P. M. Echenique, J. M. Pitarke, E. V. Chulkov, and V. M. Silkin, Image-potential-induced states at metal surfaces, J. Electron Spectrosc. Relat. Phenom. **126**, 163 (2002).

[70] N. Lu and J. Yang, Electronic structure engineering via on-plane chemical functionalization: a comparison study on two-dimensional polysilan and graphane, J. Phys. Chem. C **113**, 16741 (2009).

[71] E. R. Margine and V. H. Crespi, Universal behavior of nearly free electron states in carbon nanotubes, Phys. Rev. Lett. 96, 196803 (2006).c

[72] T. Miyake and S. Saito, Electronic structure of potassium-doped carbon nanotubes, Phys. Rev. B 65, 165419 (2002)].

[73] J. Guo, Q. Peng, H. Fu, G. Zou, and Q. Zhang, Heavy-metal adsorption behavior of two-dimensional alkalization-intercalated MXene by first-principles calculations, J. Phys. Chem. C **119**, 20923 (2015).

[74] T. Inoshita, S. Jeong, N. Hamada, and H. Hosono, Exploration for two-dimensional electrides via database screening and *ab initio* calculation, Phys. Rev. X **4**, 031023 (2014).

[75] S. *J.* Grimme, Semiempirical GGA-type density functional constructed with a long-range dispersion correction, Comput. Chem. **27**, 1787 (2006).

[76] D. Er, J. Li, M. Naguib, Y. Gogotsi, and V. B. Shenoy, $T_3C_2$ MXene as a high capacity electrode material for metal (Li, Na, K, Ca) ion batteries, ACS Appl. Mater. Interfaces **6**, 11173 (2014).

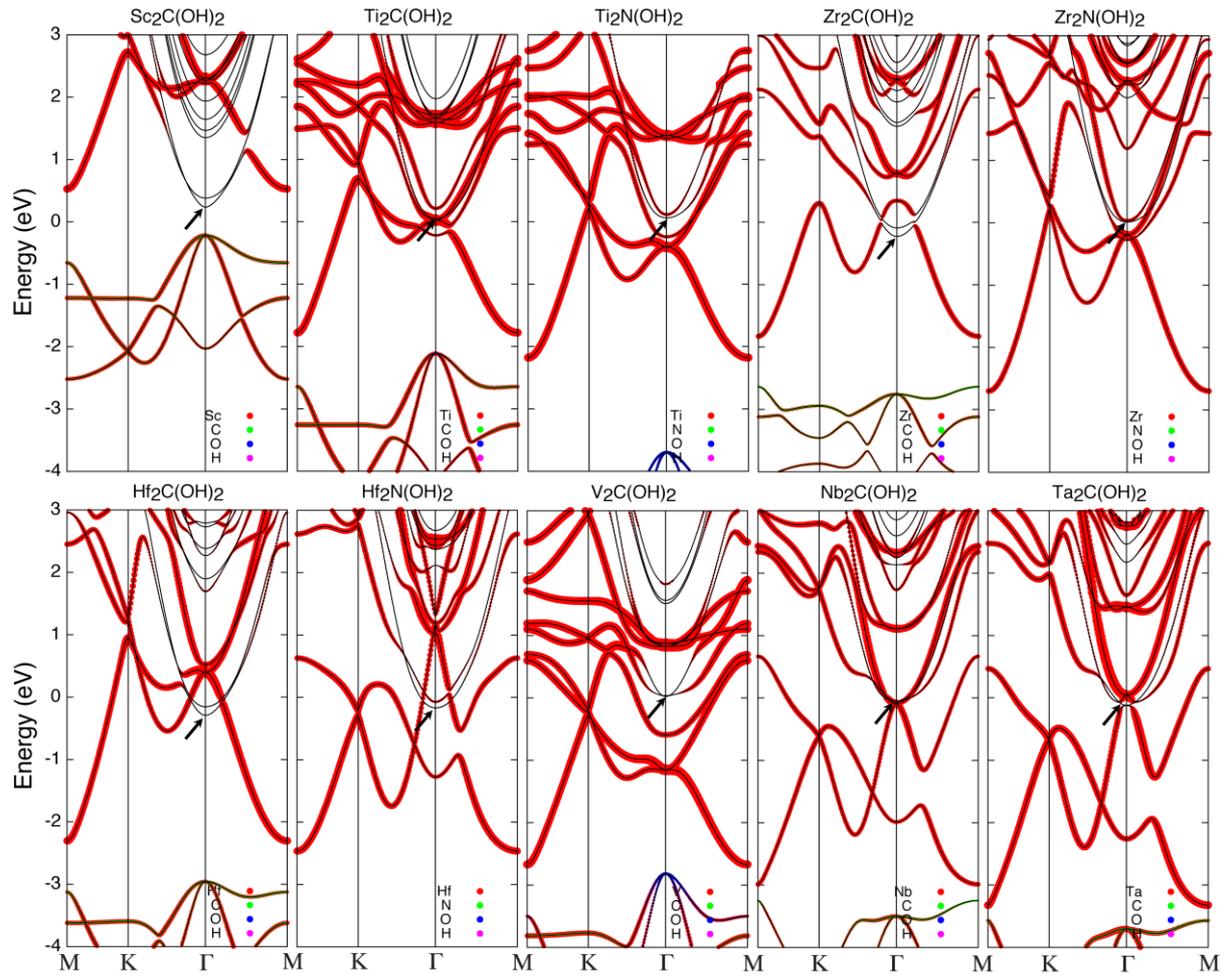

**FIG. 1**. Projected energy band structures of various OH terminated MXenes. The larger diameter of the dots indicates the larger contribution of the constituent elements to the wave functions. The Fermi energy is located at zero. The arrows indicate the lowest energy NFE state near the Fermi energy.

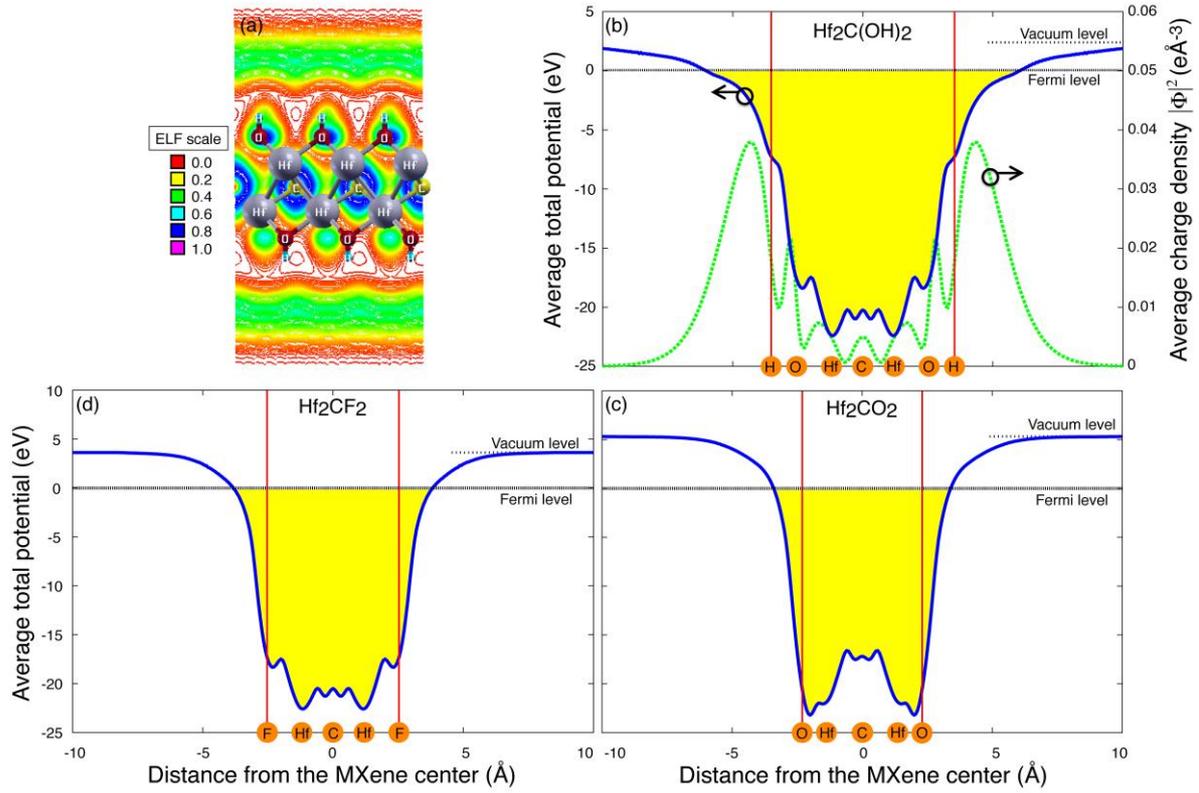

**FIG. 2**. (a) Electron localization function (ELF) contour plot for $Hf_2C(OH)_2$. (b)-(d) Averaged total potential (blue solid lines) along the axis perpendicular to the surfaces of (b) $Hf_2C(OH)_2$, (c) $Hf_2CO_2$, and (d) $Hf_2CF_2$. The vertical lines indicate the position of the outmost surface atoms. The local potential regions where the averaged total potential is lower than the Fermi level are indicated in yellow color. In (b) averaged charge density for the lowest NEF state at Γ point for $Hf_2C(OH)_2$ is also shown in green dotted line. Note that the vacuum space of 40 Å thick is taken in the calculations, although the figure is plotted only for the region near the center of the MXenes.

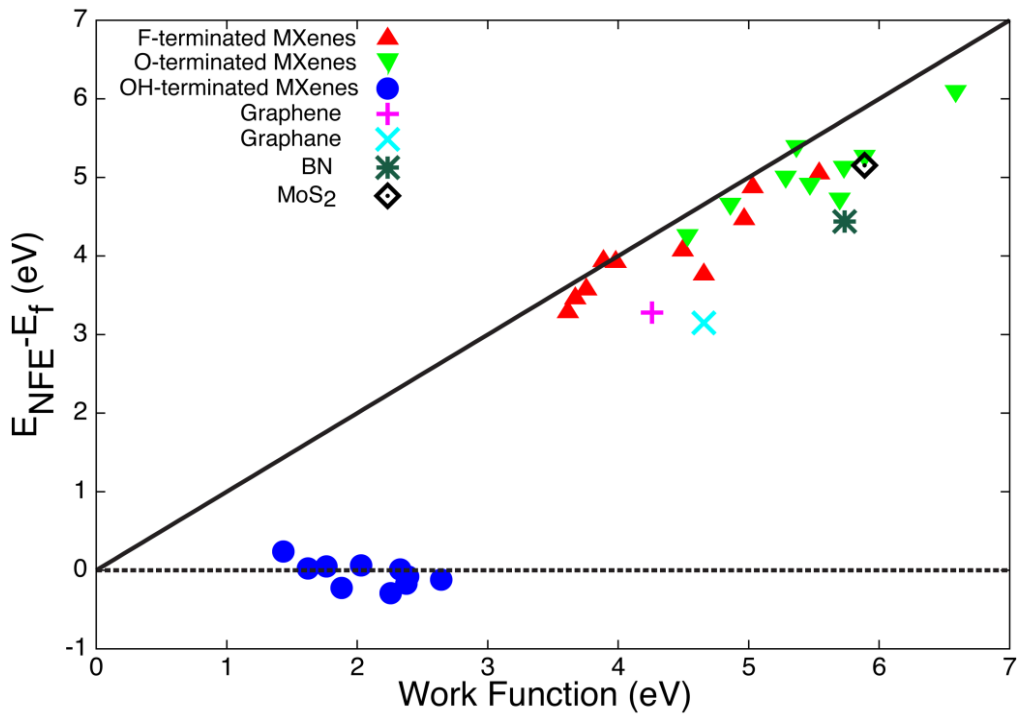

**FIG. 3.** The relative energy position of the lowest NFE state for functionalized MXenes, as well as for graphene, graphane, BN, and $MoS_2$ layers with respect to the Fermi level ($E_f$) as a function of the work function. The solid line represents the energy position of the vacuum level.

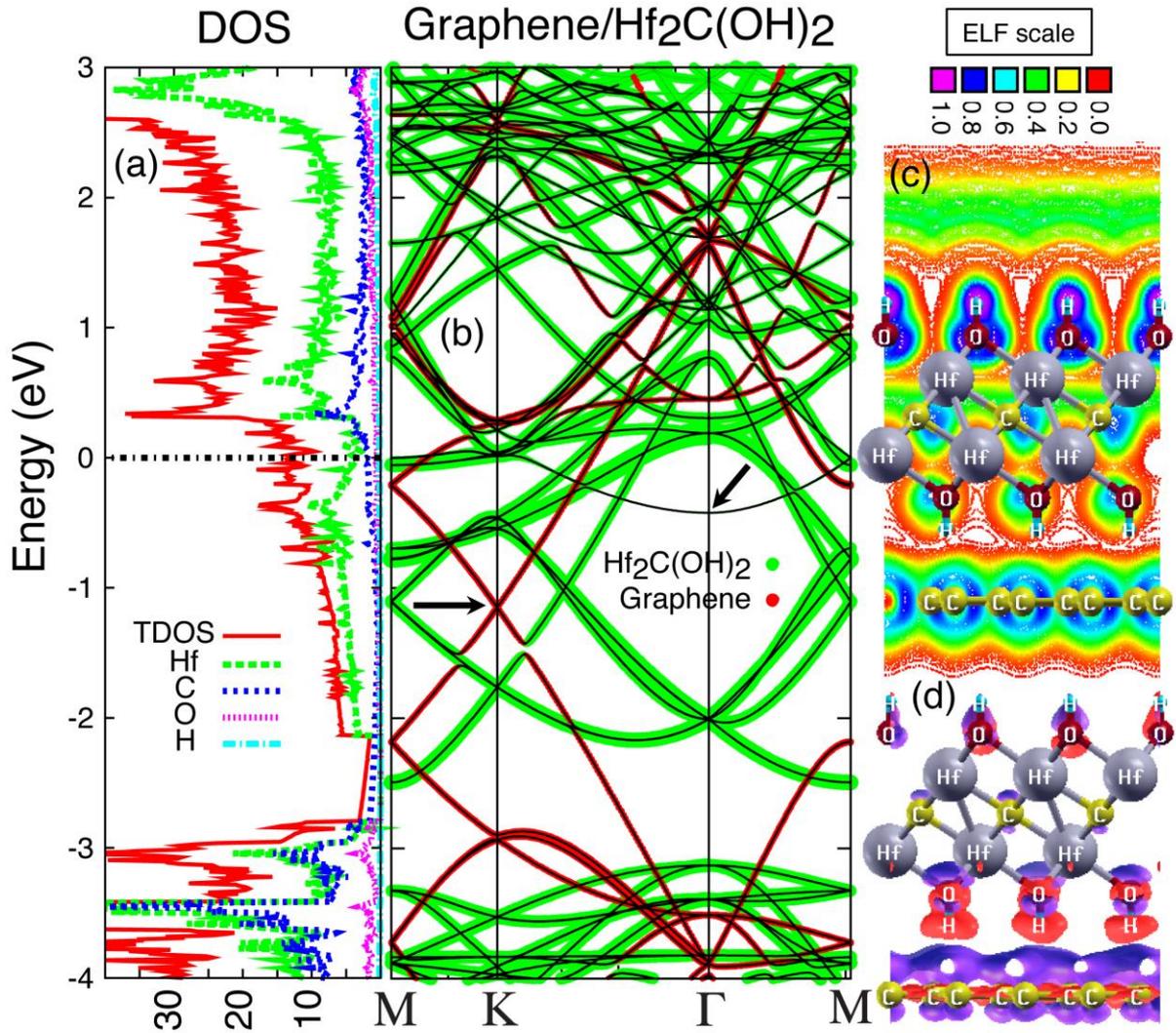

**FIG. 4**. (a) and (b) Density of states (DOS in states/eV/cell) and projected band structure of graphene/$Hf_2C(OH)_2$ heterostructure. Arrows in (b) indicate the position of the Dirac cone at K and the NFE band at Γ. The Fermi level is located at zero. (c) Electron localization function (ELF) counter plot. (d) The charge transfer (Δρ) isosurfaces for ±0.005 e/Å$^3$. The excess and depleted charges are shown in dark orchid and deep pink colors.

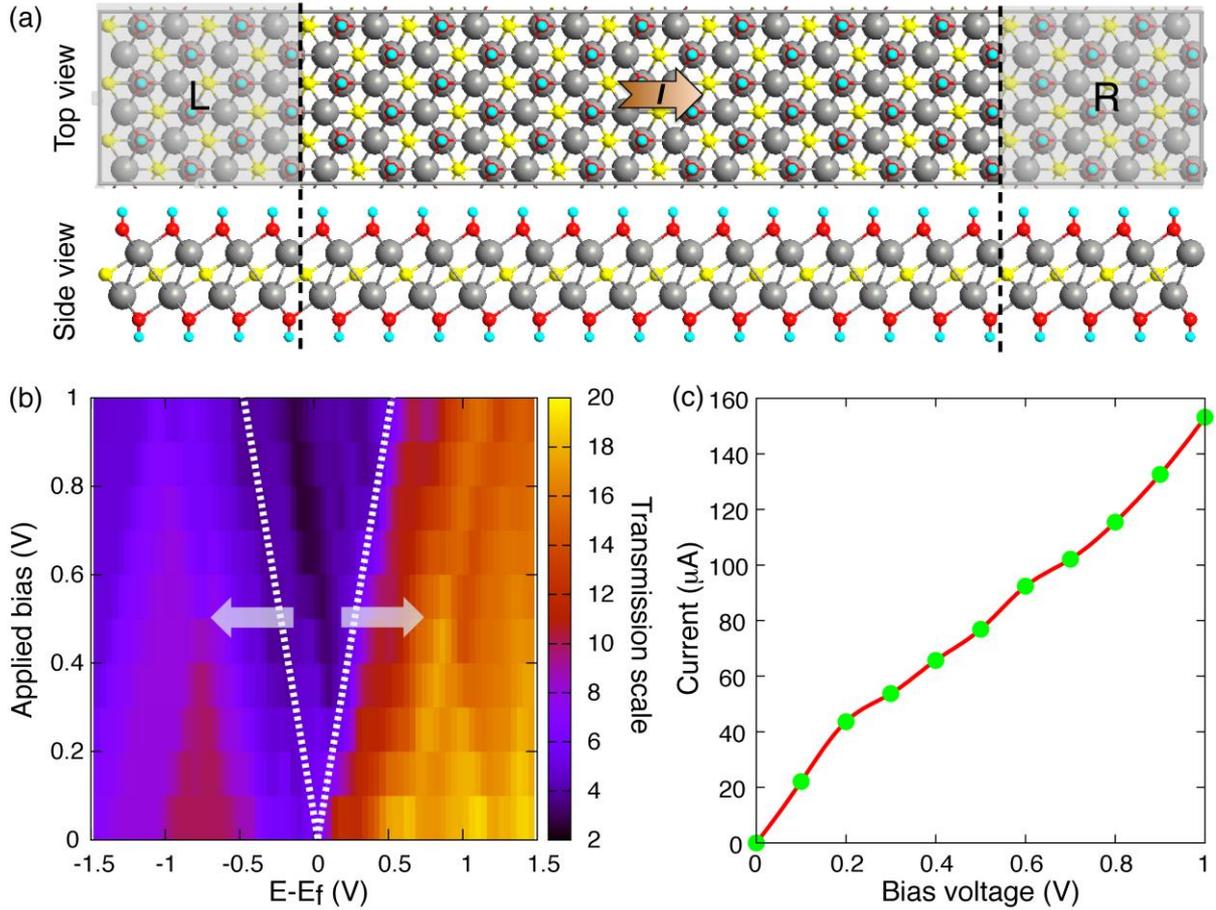

**FIG. 5**. (a) Schematic representation of the setup for the electronic transport calculation of $Hf_2C(OH)_2$. The length and width of the scattering region (the central region encompassed by the dashed lines) are 40.100 Å and 9.926 Å, respectively. The highlighted regions by L and R symbols represent the semi-infinite left and right electrodes. (b) Transmission spectrum at different bias voltages. The Fermi level is indicated as $E_f$. (c) current–voltage characteristics for $Hf_2C(OH)_2$. The Fermi level ($E_f$) locates at zero energy. The white dotted lines in (b) indicate the integration region for current calculations. The white arrows in (b) indicate the shift of transmission peaks for electrons and holes as the bias voltage increases.

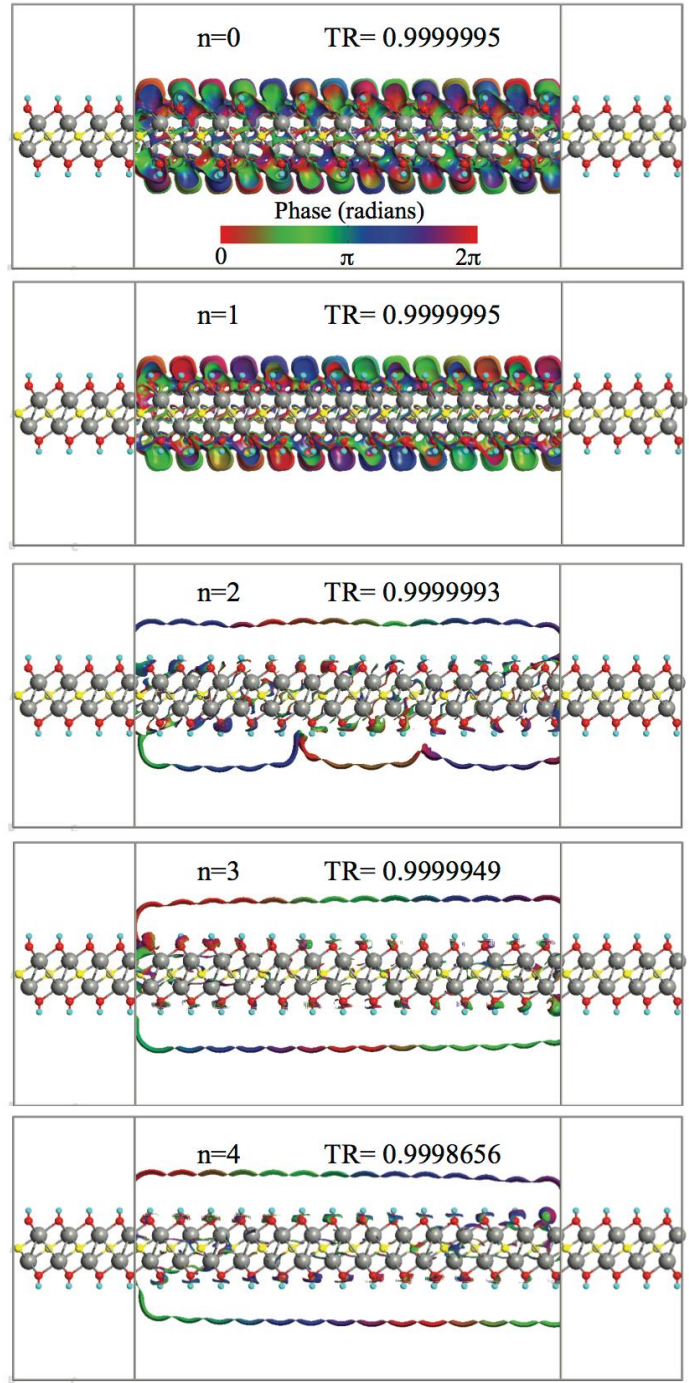

**FIG. 6**. Side views of the eigenvectors of transmission matrix at k=(0,0) and energy -0.088 eV at zero bias voltage. TR indicates the eigenvalue of transmission matrix. Since the transmission eigenvector is complex, the absolute value of the eigenvector is shown by the isosurface and the phase is indicated by the color of the isosurface. The transmission channels n = 3 and 4 exhibit the NFE characteristics.